\documentclass{article}
\usepackage{spconf,amsmath,graphicx}
\usepackage{amsfonts} 
\usepackage{multirow}
\usepackage{mathtools}
\usepackage{booktabs}
\usepackage{diagbox}
\usepackage[table,x11names,dvipsnames]{xcolor}
\usepackage[breaklinks,colorlinks]{hyperref}
\usepackage{caption}

\usepackage{etoolbox}
\apptocmd{\thebibliography}{\setlength{\itemsep}{-2pt}}{}{}
\patchcmd{\thebibliography}
  {\settowidth}
  {\setlength{\itemsep}{0pt plus 0.0pt}\settowidth}
  {}{}
\apptocmd{\thebibliography}
  {\small}
  {}{}

\definecolor{LightGrey}{rgb}{0.9,0.9,0.9}

\newcommand\tstrut{\rule{0pt}{2.2ex}}
\newcommand\bstrut{\rule[-1.0ex]{0pt}{0pt}}

\title{Play It Back: Iterative Attention for Audio Recognition}
\name{\normalsize Alexandros Stergiou$^{1,2,*}$, Dima Damen$^3$}

\address{
\noindent $^1$Vrije University of Brussels, Belgium $\quad$
$^2$Interuniversity Microelectronics Centre, Leuven, Belgium\\
$^3$University of Bristol, United Kingdom}
\begin{document}
%
\maketitle

{\let\thefootnote\relax\footnote{{$\!^*$Work was done while A. Stergiou was at the University of Bristol.}}}
\setcounter{footnote}{0}

\vspace*{-20pt}
\begin{abstract}
A key function of auditory cognition is the association of characteristic sounds with their corresponding semantics over time.
Humans attempting to discriminate between fine-grained audio categories, often replay the same discriminative sounds to increase their prediction confidence.
We propose an end-to-end attention-based architecture that through selective repetition attends over the most discriminative sounds across the audio sequence. Our model initially uses the full audio sequence and iteratively refines the temporal segments replayed based on slot attention. At each playback, the selected segments are replayed using a smaller hop length which represents higher resolution features within these segments. 
We show that our method can consistently achieve state-of-the-art performance across three audio-classification benchmarks: AudioSet, VGG-Sound, and EPIC-KITCHENS-100. \protect\footnote{Our code is available at: \url{tinyurl.com/playitback2023}}
\end{abstract}
%

\section{Introduction}
\label{sec:intro}

Audio recognition is the task of categorizing audio with discrete labels that semantically represent the emitted sounds. This includes significant challenges considering the similarity in object sounds (e.g. boat motors and road vehicles), musical instruments (e.g. guitar, banjo, and ukulele), human (e.g. wail and groan), or animal (e.g. yip and growl) sounds. 


In everyday life, we repeat parts of songs or ask for someone to repeat themselves to better understand audio. This relates to the development of echoic memory which is responsible for the memorization of sounds \cite{clark1987echoic,strous1995auditory}. Therefore, repeated listens and replays of sound stimulants \cite{radvansky2005human} are an essential part of learning and associating sound patterns.  

\begin{figure}[ht]
    \centering
    \includegraphics[width=\linewidth]{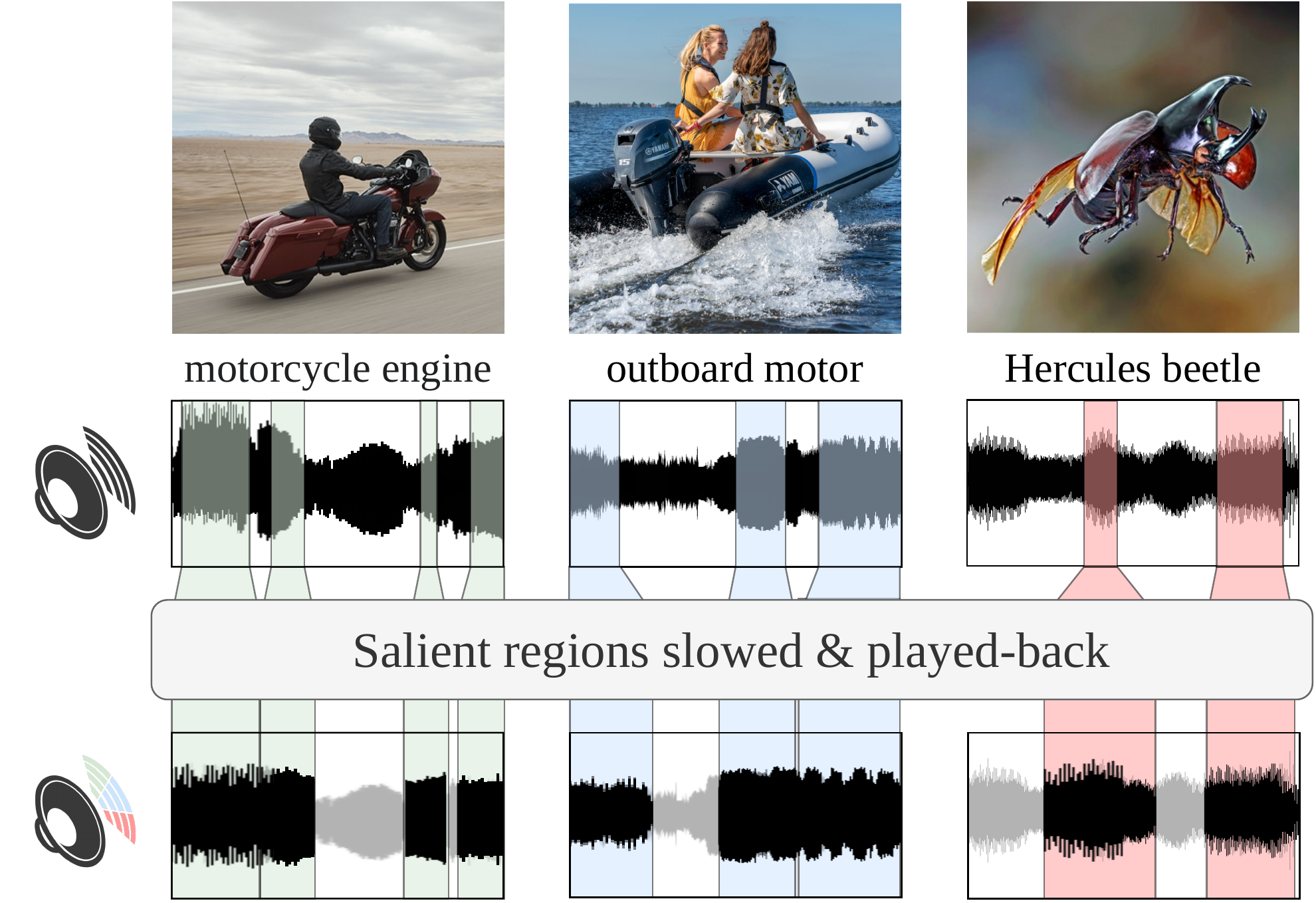}
    \caption{\textbf{Playback of discriminative sounds}. Given an audio sequence, the most relevant sounds are selected and played back at reduced hop length. The generated playbacks attend solely informative sounds at a higher temporal resolution.}
    \vspace*{-12pt}
    \label{fig:cover}
\end{figure}

Driven by the perception of sound through echoic memory and the recent success of Vision Transformers (ViT) \cite{dosovitskiy2020image} at utilizing global context information, we propose an end-to-end attention-based model that recognizes sounds through discovering and playing back the most informative sounds from the audio sequence, as shown in Figure~\ref{fig:cover}. We use slots \cite{locatello2020object}
to attend to category-relevant sounds in the input sequence.
These slots select the time segments to be replayed. Coarser features from earlier playbacks are memorized alongside finer (i.e. higher-temporal resolution) features from later playbacks with the use of a transformer decoder.

Our contributions are as follows: i) We propose to select and replay relevant audio features with decreased hop lengths, slowing down relevant parts of the audio. ii) We propose an end-to-end transformer architecture for audio recognition that jointly selects and attends to multiple audio replays, and refines the final class predictions. iii) Our method achieves state-of-the-art performance on AudioSet \cite{gemmeke2017audio}, VGG-Sound~\cite{chen2020vggsound}, and EPIC-KITCHENS-100 \cite{damen2022rescaling}.

\begin{figure*}[ht]
\centering
\includegraphics[width=\linewidth]{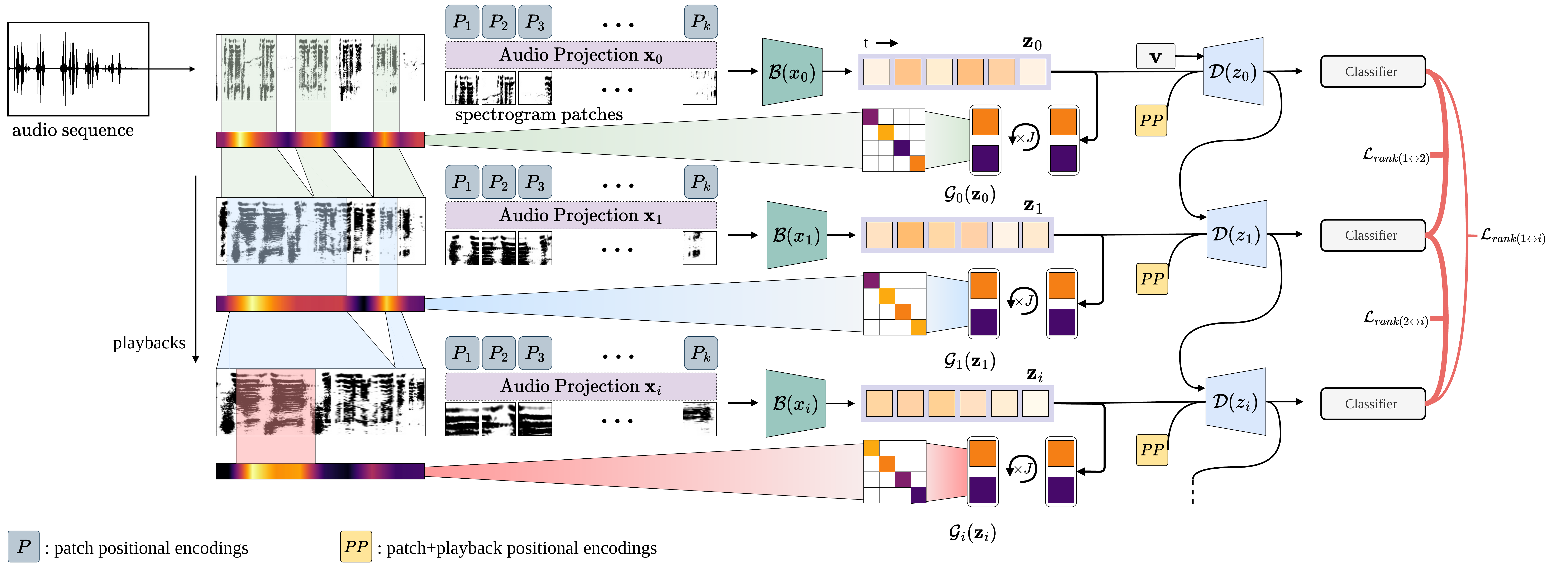}
\caption{\textbf{PlayItBack architecture}. 
The spectrogram of the full audio sequence (top) is replayed by focusing on discriminative features and reducing the hop length to capture finer temporal details (bottom).
During each playback, spectrogram patches are tokenized~$\mathbf{x}_i$ and appended patch (frequency and temporal) positional encodings ($P$). Several multi-head attention layers are used to encode features $\mathbf{z}_i$. Slot attention $\mathcal{G}(\mathbf{z}_i)$ then discovers discriminative temporal segments. 
These are considered input to the next playback.
To combine decisions between playbacks, a recurrent Transformer Decoder $\mathcal{D}(\mathbf{z}_i)$ takes previously decoded features from $i$ and the encoded features in $i+1$ playback appended with patch encodings ($PP$). PlayItBack is trained by classification loss $\mathcal{L}_{CLS_i}$, regularized by the weighted sum of ranking losses $\mathcal{L}_{rank_i}$ between $i$ and $\{1,...,i-1\}$ playbacks.}
\vspace*{-10pt}
\label{fig:playitback}
\end{figure*}

\section{Related Work}
\label{sec:related}

\noindent
\textbf{Audio recognition}. A popular approach for audio classification has been the use of convolutional networks, previously used for image-based object recognition \cite{gong2021psla,kong2020panns,wang2019comparison} or video classification \cite{kazakos2021slow} tasks, to learn features from audio spectrograms. The introduction of Transformer-based architectures has further given rise to their adaptation for audio recognition by works relying on hybrid architectures \cite{kong2020sound,gulati2020conformer,miyazaki2020convolution}. Similar attempts have also built on image-pretrained Transformer models for attending audio spectrograms \cite{chen2022hts,koutini2021efficient}. \cite{nagrani2021attention} incorporated an additional video modality to improve performance. 
Recently, \cite{liu2022learning} have also studied the effects of different hop lengths on the temporal resolution of spectrograms.

In contrast to the majority of previous works, we focus on identifying relevant and irrelevant sounds. The irrelevant sounds are removed, while relevant segments are slowed and replayed with predictions calculated across \textit{playbacks}. 



\noindent
\textbf{Selecting discriminative features}. Modeling discriminative features has been a central focus of image recognition methods. \cite{wang2017autoscaler} generated features from multiple scales selecting the best-suited features per scale. \cite{rosenfeld2016visual} proposed the aggregation of features from image regions cropped based on class saliencies. \cite{recasens2018learning} applied distortion grids on the cropped regions. Most similar to our work, \cite{fu2017look} used a recurrent CNN to select image regions that are attended to in follow-up scales. In their work, only a single region was selected per scale. 
Instead, we identify multiple discriminative sound segments which we combine to form the next playback.

We describe our PlayItBack method next.



\section{Method}
\label{sec:pagestyle}

In this section, we describe the proposed PlayItBack architecture, depicted in Figure~\ref{fig:playitback}. We compute the mel-spectrogram for a given audio sequence resulting in an $F \! \times \! T$ representation of frequency $F$ and time $T$, and extract $k$ non-overlapping patches. We project the patches to feature tokens $\mathbf{x}_i \!\! \in \!\! \mathbb{R}^{D}$, where $D \! = \! FT$. A transformer encoder $\mathcal{B}$ is used to encode these into features $\mathbf{z}_{i}$. Slot attention $\mathcal{G}$ is applied to $\mathbf{z}_{i}$ to select the discriminative regions which, at the next playback, will be slowed by decreasing the hop length in the spectrogram. The Decoder $\mathcal{D}$ then relates features across playbacks.


\noindent
\textbf{Transformer Encoder}. Given linear projections $\mathbf{x}_i$, we use frequency and temporal patch positional encodings $P$. The encoder network $\mathcal{B}$ extracts representations for each playback, $\mathbf{z}_i \! = \! \mathcal{B}(\mathbf{x}_i) \! \in \! \mathbb{R}^{d \times C}, \text{with} \; d \! < \!\! FT$ resolution and $C$ channels.

\begin{table*}[t]
\parbox{.49\linewidth}{
\centering
\resizebox{.98\linewidth}{!}{
\begin{tabular}{l  l  l c }
\hline
Model & Backbone & Train set & mAP \\
\hline
\rowcolor{LightGrey} \multicolumn{4}{l}{\textit{Audio-only models}} \tstrut \\
MAE-AST \cite{baade2022mae} & ViT-B \cite{dosovitskiy2020image} & mini-AS & 30.6\\
Perceiver \cite{jaegle2021perceiver} & Perceiver & AS-2M & 38.4 \\
Conformer \cite{gulati2020conformer} & Conformer & AS-2M & 41.1 \\
PANN \cite{kong2020panns} & ResNet38 \cite{he2016deep} & AS-2M & 43.4 \\
MBT \cite{nagrani2021attention} & ViT-B & AS-500K & 44.3\\
PSLA \cite{gong2021psla} & EffNet-B2 \cite{tan2019efficientnet} & AS-2M & 44.4 \\
PaSST \cite{koutini2021efficient} & DeiT-B \cite{touvron2021training} & AS-2M & 47.1\\
HTS-AT \cite{chen2022hts} & Swin-T \cite{liu2021swin} & AS-2M & 47.1 \\
MaskSpec \cite{chong2022masked} & ViT-B & AS-2M & 47.1 \\
Audio-MAE \cite{xu2022masked} & ViT-B  & AS-2M & 47.3 \\
\hline
\textbf{PlayItBackX3} & MViTv2-B \cite{li2022mvitv2} & AS-500K & \textbf{47.7} \\
\end{tabular}
}
\caption{\textbf{Comparisons to state-of-the-art audio-only models on AudioSet}. We report the mean average precision (mAP) alongside the backbone and training set used.\label{tab:audioset}}
}
\hfill
\parbox{.49\linewidth}{
\centering
\resizebox{.98\linewidth}{!}{
\begin{tabular}{l l l c c c }
\hline
Model & top-1 & top-5 & mAP & AUC & d-prime \tstrut \bstrut \\
\hline
\rowcolor{LightGrey} \multicolumn{6}{l}{\textit{Audio-only models}} \tstrut\\[.2em]
McDonnell \& Gao \cite{mcdonnell2020acoustic} & 39.7 & 71.6 & 40.3 & 0.963 & 2.532 \\[.15em]
Peng et al. (A) \cite{peng2022balanced} & 44.3 & - & 48.4 & - & - \\[.15em]
ResNet-101 \cite{kazakos2021slow} & 45.6 & 72.3 & 47.6 & 0.968 & 2.615 \\[.15em]
Chen et al. \cite{chen2020vggsound} & 51.0 & 76.4 & 53.2 & 0.973 & 2.735 \\[.15em]
MBT (A) \cite{nagrani2021attention} & 52.3 & 78.1 & - & - & - \\[.15em]
Slow-Fast \cite{kazakos2021slow} & 52.4 & 78.1 & 54.4 & 0.974 & 2.761 \\[.15em]
\hline
\textbf{PlayItBackX3} & \textbf{53.7} & \textbf{79.2} & \textbf{56.1} & \textbf{0.978} & \textbf{2.846} \tstrut \bstrut \\
\hline
\rowcolor{LightGrey} \multicolumn{6}{l}{\textit{Models trained with additional modalities}} \tstrut \\[.2em]
\textcolor{gray}{Peng et al. (AV) \cite{peng2022balanced}} & \textcolor{gray}{50.6} & - & \textcolor{gray}{53.9} & - & - \\[.15em]
\textcolor{gray}{PolyViT \cite{likhosherstov2021polyvit}} & \textcolor{gray}{51.7} & - & - & - & - \\[.15em]
\textcolor{gray}{MBT (AV)} & \textcolor{gray}{64.1} & \textcolor{gray}{85.6} & - & - & - \\
\end{tabular}
}
\caption{\textbf{Comparisons to state-of-the-art models on VGG-Sound}. We report the top-1 and top-5 accuracies (\%) alongside mAP, the AUC and d-prime.\label{tab:vggsound}}
}
\hfill
\vspace{-0.5em}
\end{table*}

\noindent
\textbf{Slot attention}. We use slot attention \cite{locatello2020object} $\mathcal{G}$ to iteratively  map the resulting feature vectors $\mathbf{z}_i$ from each playback to two slot vectors $\mathbf{s}_{lj}$ corresponding to the informative $\mathbf{s}_{1j}$ and uninformative $\mathbf{s}_{2j}$ temporal segments of the audio input respectively. We use $j \! \in \! \{1,...,J\}$ to denote slot iterations. The query $\mathbf{Q}_{lj} \! = \! MLP(LN(\textbf{s}_{lj-1}))$, key $\mathbf{K}_{lj} \! = \! MLP(LN(\textbf{z}_i))$ and value $\mathbf{V}_{lj} \! = \! MLP(LN(\textbf{z}_i))$ use Layer Normalization $LN(\cdot)$ followed by Multi-Layer Perceptron $MLP(\cdot)$ to map the features $\mathbf{z}_i$ and slots $\mathbf{s}_{lj}$ vectors to a common dimension $d$. We set the softmax temperature based on a fixed value $\sqrt{d}$. 
\begin{equation}
\begin{aligned}
\label{eq:slot_attn}
    \mathbf{h}_{lj} = GRU \Bigl(\frac{\mathbf{a}_{lj} \: \mathbf{V}_{lj}}{\smashoperator{\sum_{m \in \{1,2\}}} \mathbf{a}_{mj}}\Bigr), \, \text{where} \; \mathbf{a}_{lj} \! = \! Attn \Bigl(\frac{\mathbf{K}_{lj} \, \mathbf{Q}_{lj}^{T}}{\sqrt{d}}\Bigr)
\end{aligned}
\end{equation}

\noindent
A Gated Recurrent Unit (GRU) with two hidden units is used at each slot iteration updating the slot hidden states $\mathbf{h}_{lj}$ as in~\cite{locatello2020object}. A linear transformation alongside a residual connection is used for the slots $\mathbf{s}_{lj} \! = \! \mathbf{s}_{lj-1} \! + \! MLP(LN(\mathbf{h}_{lj}))$. 

We train the two slots so $\mathbf{s}_1$ attends to informative audio while $\mathbf{s}_2$ captures the remaining audio.
This is achieved by combining $\mathcal{G}(\mathbf{z}_{i})_{1}\! = \! \mathbf{s}_{1}$ and the \emph{inverse} of the uninformative slot $\mathcal{G}(\mathbf{z}_{i})_{2} \! = \! \mathbf{s}_{2}$ to create the attention matrix: ${\mathbf{M} \! = \! Attn(\mathcal{G}(\mathbf{z}_{i})_{1}^{T} \, \mathcal{G}(\mathbf{z}_{i})_{2}^{-1})}$. We normalize and rescale the main diagonal $diag(\mathbf{M})$ by interpolation so that it matches the temporal dimension of $\mathbf{x}_i$. Activations above the normalized average ($> \! 0.5$) are selected for the segments in~$\mathbf{x}_{i+1}$.


\noindent
\textbf{Transformer Decoder}. Given the extracted encoder features~$\mathbf{z}_i$, the decoder transformer $\mathcal{D}$ relates information across playbacks. Positional encodings based on patches and the playback number are added to~$\mathbf{z}_i$. Considering the iterative nature of the PlayItBack model, cross-attending~\cite{jaegle2021perceiver} information over playbacks enables the model to retain general features and associate patterns that are common. For the decoder, 
we define the query from the previous playback as $\mathbf{Q}_{i} \! = \! MLP(LN(\mathbf{v}_i))$, where $\textbf{v}_1$ is initialized with a latent vector then updated at each playback $\textbf{v}_{i} = \mathcal{D}(\textbf{z}_{i-1}, \textbf{v}_{i-1})$, where $i > 1,$ 
key $\mathbf{K}_{i} = MLP(LN(\textbf{z}_i))$ and value $\mathbf{V}_{i} = MLP(LN(\textbf{z}_i))$ for the cross attention. This is followed by a self-attention block. The decoder features are then passed to a classifier shared across playbacks.


\noindent
\textbf{Classification and rank loss}. We use an inter-playback weighted ranking loss $\mathcal{L}_{\text{rank}(i)}$ for forcing the network to attain more confident predictions in later playbacks. The ranking loss $ \mathcal{L}_{\text{rank}(i)}$, uses the pair-wise class probabilities $p(\omega)_i$ and $p(\omega)_{m} \forall m \! \in \! \{1,...,i\!-\!1\}$ for the correct class label $\omega$. We compute the probability difference $\mathcal{L}_{\text{rank}(m \leftrightarrow i)}$ between the $i$th playback and all previous playbacks.
\begin{equation}
    \mathcal{L}_{\text{rank}(i)} \! = \! \smashoperator{\sum_{m = 1}^{i-1}} \lambda_m \, 
    max(0, \gamma - p(\omega)_{i} + p(\omega)_{m})
\end{equation}
The ranking loss thus uses predictions from the previous playbacks as a reference with the expectation that $p(\omega)_i \! > \! p(\omega)_m + \gamma$, i.e. subsequent playbacks always increase confidence, where $\gamma$ is the ranking loss's margin. For stability in training, we include a weight $\lambda_m = \frac{1}{i-m}$ computed based on the difference between the playback indices. 

We combine $\mathcal{L}_{rank(i)}$ with an inter-playback cross-entropy loss $\mathcal{L}_{\text{CLS}(i)}$ and define our multi-task loss as: 
\begin{equation}
\label{eq:loss}
   \mathcal{L} = \mathcal{L}_{\text{CLS}(1)} + \smashoperator{\sum_{i=2}^N} \mathbf{\beta} \, \mathcal{L}_{\text{CLS}(i)} + (1-\mathbf{\beta}) \, \mathcal{L}_{\text{rank}(i)}
\end{equation}
where $\mathbf{\beta}$ is a weighting parameter for the aggregation of the cross-entropy and ranking losses. During inference, our model uses the average of all predictions across playbacks.  



\section{Experiments}
\label{sec:experiments}

\noindent
\textbf{Datasets} We evaluate our proposed PlayItBack architecture on three large-scale datasets. \textbf{AudioSet} \cite{gemmeke2017audio} is composed of 2M 10s audio clips from YouTube annotated with 527 classes (AS-2M). Because of the high imbalance of the dataset, we instead train with the proposed AS-500K~\cite{nagrani2021attention}.  \textbf{VGG-Sound}~\cite{chen2020vggsound} consists of 200k clips of 10s length with 309 labels corresponding to human actions, objects and interactions. \textbf{EPIC-KITCHENS-100} \cite{damen2022rescaling} includes 90k clips of hand-object interactions labeled with 97 verb, 300 noun classes, and 4025 action classes. The clip length is variable and 2.6s on average.

\noindent
\textbf{Evaluation metrics}. For AudioSet along the lines of previous works, we use the mean average precision (mAP). For VGG-Sound, as in \cite{kazakos2021slow}, we report the top-1/5 \% accuracies, mAP, AUC, and d-prime. For EPIC-KITCHENS-100 we report the top-1/5 \% accuracies for the verb, noun, and action labels.  

\noindent
\textbf{Implementation details}. We use PlayItBackX3 with $N\!\!=\!\!3$ as our model for comparative evaluation, with ablations showcasing that this produces the best accuracy (top-1)/compute (GFLOPs) trade-off. We use the 24-layer MViTv2-B \cite{li2022mvitv2} as our default encoder\footnote{The flattened vector size is d=50 and the number of features is C=768}. We note that due to the fixed number of 2D patches used by MViTv2, the spectrogram dimensions remain constant throughout playbacks. For all experiments, we set the ranking margin $\gamma\!\!=\!\!0.05$, $J=3$ slot iterations, and $\mathbf{\beta}\!=\!0.7$. As in \cite{kazakos2021slow} we use spectrograms with frequency dimension of 128 corresponding to inputs of size $128 \!\! \times \!\! 100S$ for $S$ seconds of audio. Our initial spectrograms are created based on the same hop length of 10ms, and 16kHz frequency as in~\cite{kazakos2021slow,nagrani2021attention,xu2022masked}.
For subsequent iterations, we reduce the hop length by 1ms at each iteration. 

We train for 50 epochs with  Mixup~\cite{zhang2018mixup} ($\alpha = 0.3$) and base learning rate of 0.5 for AudioSet and 0.01 for VGG-Sound \& EPIC-KITCHENS-100. We use warm-up for the first 2.5 epochs, a decayed cosine schedule, batch size of 64 with SGD, momentum set to 0.9, and $1e^{-4}$ weight decay.

\begin{table}[t]
\centering
\resizebox{\linewidth}{!}{
\begin{tabular}{l c c c c c c c }
\hline
\multirow{2}{*}{Model} & \multirow{2}{*}{GFLOPs} & \multicolumn{2}{c}{verb} & \multicolumn{2}{c}{noun} & \multicolumn{2}{c}{action} \tstrut \bstrut \\ \cmidrule(rl){3-4} \cmidrule(rl){5-6} \cmidrule(rl){7-8}
& & top-1 & top-5 & top-1 & top-5 & top-1 & top5 \tstrut \bstrut \\
\hline
Damen et al. \cite{damen2022rescaling} & N/A & 42.6 & 75.8 & 22.3 & 44.6 & 14.5 & 28.2 \tstrut \\
MBT (A) \cite{nagrani2021attention} & 34.2 & 44.3 & - & 22.4 & - & 13.0 & - \tstrut \bstrut \\[.1em]
Slow-Fast \cite{kazakos2021slow} & 35.1 & 46.5 & 78.3 & 22.8 & 44.9 & 15.4 & 28.6 \bstrut \\[.1em]
\hline
\textbf{PlayItBackX3} & 122.8 & \textbf{47.0} & \textbf{78.7} & \textbf{23.1} & \textbf{45.1} & \textbf{15.9} & \textbf{29.2} \tstrut \\
\end{tabular}
}
\vspace{-.5em}
\caption{\textbf{Comparisons to state-of-the-art for EPIC-KITCHENS-100}. We report the top-1 and top-5 accuracies for the verb, noun, and action labels.}
\label{tab:epic}
\vspace{-.8em}
\end{table}

\begin{table}[t]
\centering
\resizebox{\linewidth}{!}{
\begin{tabular}{l l l l c c c }
\hline
Model & freq. & top-1 & top-5 & mAP & AUC & d-prime \tstrut \bstrut \\
\hline
PlayItBackX0 & 32kHz & 52.1 & 77.8 & 54.7 & 0.970 & 2.757 \tstrut \bstrut \\
\hline
PlayItBackX0 & 16kHz & 51.8 & 77.4 & 54.3 & 0.966 & 2.743 \tstrut \\
PlayItBackX1 & 16kHz & 52.5 & 78.3 & 55.1 & 0.972 & 2.789 \\
PlayItBackX2 & 16kHz & 53.2 & 78.7 & 55.5 & 0.976 & 2.810 \\
PlayItBackX3 & 16kHz & \noindent \textbf{53.7} & \noindent \textbf{79.2} & \noindent \textbf{56.1} & \noindent \textbf{0.978} & \noindent \textbf{2.846}
\vspace{-.5em}
\end{tabular}
}
\caption{\noindent \textbf{Frequency to playbacks on VGG-Sound} given top-1 and top-5 accuracies, mAP, AUC, and d-prime.}
\label{tab:vggsound_freq}
\vspace{-1em}
\end{table}

\noindent
\textbf{Results}. We compare PlayItBack to current state-of-the-art models on \textbf{AudioSet} in Table~\ref{tab:audioset}. PlayItBackX3 achieves the best performance in comparison to other models. 

We report results on \textbf{VGG-Sound} in Table~\ref{tab:vggsound}. PlayItBackX3 performs favorably to in-domain audio models. Compared to the previously top-performing SlowFast model~\cite{kazakos2021slow}, we observe a +1.3\%p. top-1 accuracy improvement. Our model is only outperformed by the multi-modal (audio-visual) version of MBT (AV) \cite{nagrani2021attention}. However, PlayItBackX3 outperforms MBT (A) in the audio-only setting.

In Table~\ref{tab:epic}, we compare to audio-classification methods on \textbf{EPIC-KITCHENS-100}. We observe that the relative improvement in performance varies across datasets (higher performance gains are observed in AudioSet and VGG-Sound, while somewhat smaller on EPIC-KITCHENS-100). We believe that this is due to EPIC-KITCHENS-100 containing audio segments of 2.6s in length on average, compared to 10s durations of AudioSet and VGG-Sound.
As the segments are already shorter, they intuitively benefit less from further playbacks by focusing on discriminative regions. Even in such settings, PlayItBackX3 demonstrates a moderate but consistent performance improvement.

\begin{figure}[t]
\vspace{-.8em}
\begin{minipage}[b]{0.4\linewidth}
    \centering
     \resizebox{\linewidth}{!}{
    \begin{tabular}[ht]{lll}
    \toprule    
    $J$ & top-1 & GFLOPs \\
    \midrule    
    \noalign{\smallskip}
    1 & 53.3 & 120.4  \\
    2 & 53.5 & 121.5  \\
    3 & \textbf{53.7} & 122.8 \\
    \bottomrule 
    \end{tabular}
    }
    \captionof{table}{\textbf{Number of slot attention iterations} ($J$) with respect to the top-1 accuracy and GFLOPs.}
    \vspace{1.4em}
    \label{tab:slot_attn_iter}
\end{minipage}
\hfill
\begin{minipage}[b]{0.55\linewidth}
    \centering
    \includegraphics[width=\linewidth,trim={1cm .3cm .5cm .5cm},clip]{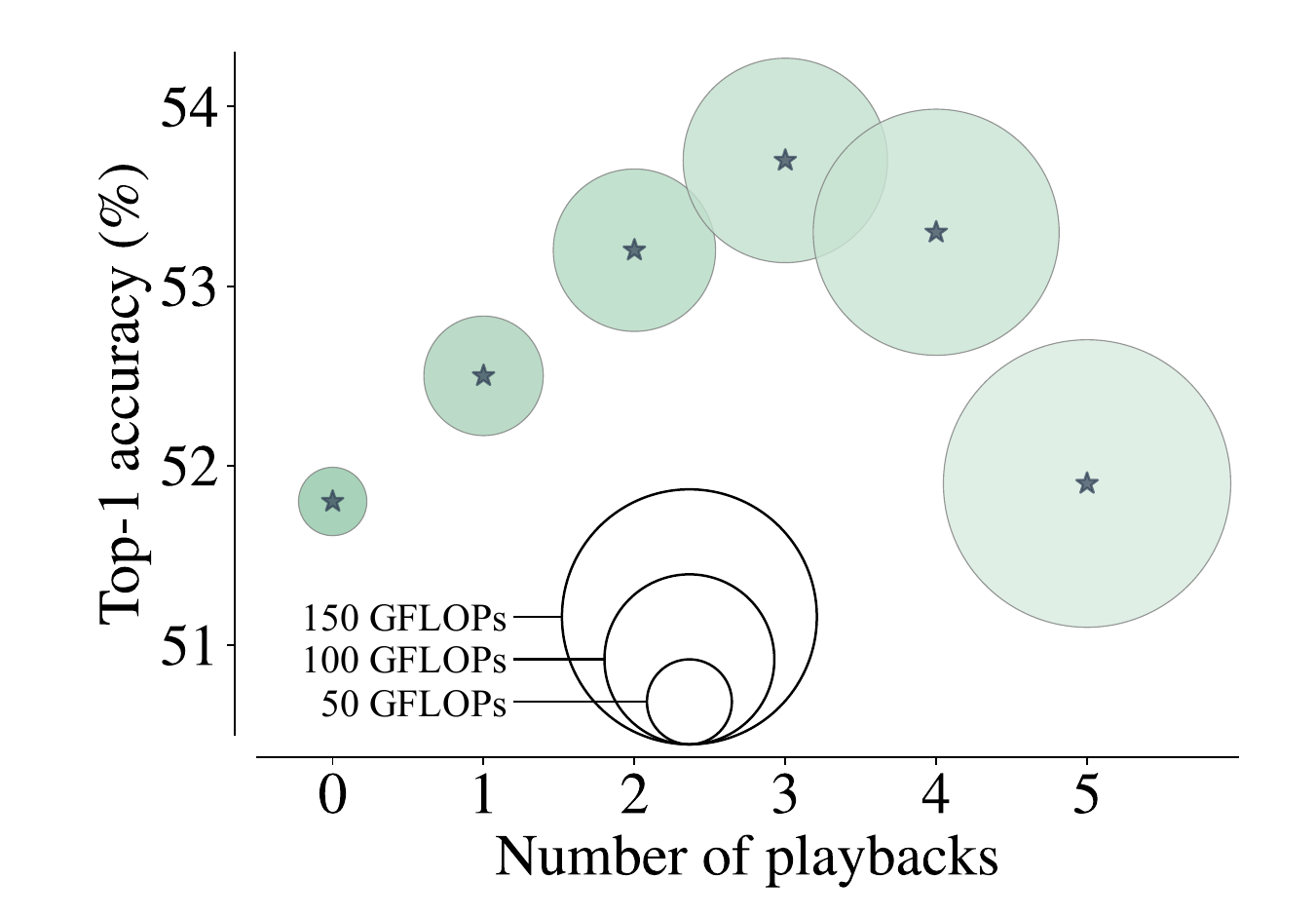}
    \captionof{figure}{\textbf{VGG-Sound top-1 accuracy over different playbacknumbers (N)} with respect to the compute (in GFLOPs). }
    \label{fig:acc2flops}
\end{minipage}
\vspace{-1em}
\end{figure}


\noindent
\textbf{Ablations}. 
Table~\ref{tab:audioset} demonstrates that while PlayItBack uses a sampling frequency of 16kHz, it can outperform HTS-AT~\cite{chen2022hts} and PaSST~\cite{koutini2021efficient} which are trained on sampling frequencies of 32kHz.
We confirm this by ablating the impact on PlayItBack.
Table~\ref{tab:vggsound_freq} demonstrates that our proposed replays at 16kHz, can be a better performing strategy than increasing the number of samples per second over the entire audio sequence as in PlayItBackX0 trained with 32kHz.

Table~\ref{tab:slot_attn_iter} compares the performance achieved with different numbers of slot iterations $J$ on VGG-Sound with PlayItBackX3. In general, moderate performance improvements can be achieved by increasing the number of slot attention iterations. The added computations also remain moderate with +2.6 GFLOPs from $J=1$ to $J=3$. In Figure~\ref{fig:acc2flops}, we investigate the impact of the number of playbacks ($N$) on the model performance. We use a decoder-only model ($N=0$), alongside PlayItBackX$N$. 
Performance improvements are shown for $1 \le N \! \le \! 3$. 
Further $N$ increases, come with performance drops, due to increased model complexity in tandem with the challenge of discovering salient information in very deep playbacks - and thus very small hop lengths.
As showcased in this figure and across results, $N=3$ offers the best performance.
The performance remains consistent over multiple runs and across datasets.

\section{Conclusions}
\vspace*{-6pt}
We propose an end-to-end attention-based architecture for audio recognition. Our PlayItBack model uses information from the full audio sequence to iteratively discover segments that are relevant to the sound. Audio segments are discovered with slot attention and amplified in the next iteration (playback). A transformer decoder is used to relate information across playbacks. 
We demonstrate the advantages of our PlayItBack approach through extensive experiments on AudioSet, VGG-Sound, and EPIC-KITCHENS-100 and ablation studies. 

Future work can explore the selection strategy for the number of playbacks, which might vary per audio sample. We hope PlayItBack can trigger insights into similar approaches for other audio signals such as speech as well as audio-visual fine-grained understanding.

\vspace*{6pt}
\noindent \textbf{Acknowledgement.} 
We use publicly available datasets.
Research is funded by the United Nation’s End Violence Fund (iCOP 2.0 project) and EPSRC UMPIRE (EP/T004991/1).

\vfill\pagebreak

\bibliographystyle{IEEEbib}
\bibliography{refs}

\end{document}